\documentclass[preprint,12pt]{aastex} 
\usepackage{natbib} 
\slugcomment{\footnotesize {\LaTeX}ed at \number\time\ min., \today} 
\shorttitle{L1221} 
\shortauthors{Young et al.} 
  
\begin{document} 

\title {The Spitzer c2d Survey of Nearby Dense Cores: VI. The Protostars of Lynds
  Dark Nebula 1221}
 
\author{Chadwick H. Young} 
 \affil{Department of Physical Sciences, Nicholls State University, 
Thibodaux, Louisiana 70310} 
 
\author{Tyler L. Bourke} 
\affil{Harvard-Smithsonian Center for Astrophysics, Cambridge, MA 02138, USA} 
 
\author{Michael M. Dunham} 
\affil{Department of Astronomy, The University of Texas at Austin, 
       Austin, Texas 78712--1083} 
 
\author{Neal J. Evans II} 
\affil{Department of Astronomy, The University of Texas at Austin, 
       Austin, Texas 78712--1083} 
 
\author{Jes K. J{\o}rgensen} 
\affil{Argelander Institut F$\ddot{u}$r Astronomie, University of Bonn, Bonn, Germany} 
 
\author{Yancy L. Shirley} 
\affil{Steward Observatory, University of Arizona, Tucson, AZ 85721} 
 
\author{Kaisa E. Young} 
\affil{Department of Physical Sciences, Nicholls State University, 
Thibodaux, Louisiana 70310} 

\author{Christopher De Vries} \affil{Department of Physics, Physical Science,
  and Geology, California State University Stanislaus, Turlock, California
  95382}

\author{Mark J. Claussen} \affil{NRAO, P.O. Box 0, 1003 Lopezville Road,
  Socorro, NM 87801}

\author{Victor Popa} 
 \affil{Department of Physical Sciences, Nicholls State University, 
Thibodaux, Louisiana 70310} 

\begin{abstract} 
  Observations of Lynds Dark Nebula 1221 from the \it Spitzer Space
  Telescope\rm\ are presented. These data show three candidate protostars
  towards L1221, only two of which were previously known.  The infrared
  observations also show signatures of outflowing material, an interpretation
  which is also supported by radio observations with the Very Large Array.  In
  addition, molecular line maps from the Five College Radio Astronomy
  Observatory are shown.
  
  One-dimensional dust continuum modelling of two of these protostars, IRS1
  and IRS3, is described.  These models show two distinctly different
  protostars forming in very similar environments.  IRS1 shows a higher
  luminosity and larger inner radius of the envelope than IRS3.  The disparity
  could be caused by a difference in age or mass, orientation of outflow
  cavities, or the impact of a binary in the IRS1 core.
 
\end{abstract} 
 
\keywords{stars: formation, low-mass, L1221} 
 
\section{Introduction} 
 
The story of star formation has long included the evolution from a starless
core through the putative series of collapse stages \citep{shu87}. These
stages include the Class 0-III, which are defined by various observational
signatures and may represent times in the life of the protostar when the
envelope and star have certain relative masses \citep{lada87}. The \it
Infrared Astronomical Satellite\rm\ (\it IRAS\rm) brought about much of what
is known about the stars forming in dense cores, and, today, the \it Spitzer
Space Telescope\rm\ reveals new light from seen and unseen stars. Through data
from \it Spitzer\rm, new details about the collapse, accretion, disk
formation, and ultimate birth of stars are being revealed.
 
In isolated, low-mass star formation, different cores provide different
laboratories to study these processes.  Some cores, such as L1014
\citep{young04}, seem to be forming only one star.  However, many isolated
cores are forming multiple star systems.  For example IRAM 04191+1522 shows
two distinct cores with 3 or more protostars \citep{andre99,dunham06}, and
IRAS 16293-2422 is forming a close binary with a third star in the system
\citep{jorgensen05,chandler05,loinard07}.  These systems are important because
they offer views of the impact that other protostars have on the evolution of
their companions.

This paper presents new infrared, spectral line, and radio continuum
observations of Lynds Dark Nebula 1221 (L1221).  Descriptions of past
observations of L1221 are in \S2, including molecular line observations and
continuum data from the infrared to millimeter. A description of the
observations and data reduction of L1221 with the \it Spitzer Space
Telescope\rm\ is in \S3.  Section 4 has results of the observations including
luminosity, classification, and other details for the infrared sources
detected by \it Spitzer\rm.  Section 5 describes the modelling of two
protostellar systems in L1221.  Finally, \S6 offers a discussion and
conclusions from the modelling.

\section{Lynds Dark Nebula 1221} 
 
Lynd's Dark Nebula 1221 (L1221) was first catalogued by \citet{lynds62}. She
listed this dark core with an area of 0.020 square degrees and a relative
opacity of 5 (on a scale of 1 to 6). The distance to L1221 is 250$\pm$50 pc
\citep{yonekura97}. \it IRAS\rm\ observed an infrared point source towards
L1221, IRAS 22266+6845. Several authors have reported on observations of
molecular line and (sub)millimeter detections.  \citet{young06} and
\citet{wu07} showed images at wavelengths of 350-850 $\mu$m, and
\citet{caselli02} reported their N$_2$H$^+$ observations of L1221.
\citet{caselli02} resolved only one extended core with a beamsize of
$1\arcmin$. However, \citet{young06} and \citet{wu07} found that L1221 has two
distinct cores, one in the north (L1221-SMM1) and one towards the south
(L1221-SMM2), these cores are separated by $50\arcsec$. L1221-SMM1 is at the
same position as IRAS 22266+6845.

Both L1221-SMM1 and SMM2 have very similar submillimeter fluxes;
\citet{young06} calculated an isothermal mass of 2.6$\pm$0.8 M$_\odot$ for
each individual core. These authors reported that L1221-SMM1, which was
detected by \it IRAS\rm, has a larger area than L1221-SMM2; \citet{wu07} gave
major and minor axes for L1221-SMM1 ($46\arcsec$ and $33\arcsec$) and
L1221-SMM2 ($28\arcsec$ and $19\arcsec$).  These axes are measured from the
2-$\sigma$ contour level.
 
Several authors have observed the outflowing material from L1221. Based on
single-dish observations, \citet{umemoto91} found a U-shaped outflow,
presumably originating from IRAS 22266+6845. They concluded that the
morphology of the outflow was due to interaction with a nearby dense ridge of
material but that some external pressure (possibly magnetic fields) was also
required to create the outflow of material seen from IRAS 22266+6845.
\citet{lee02} reported, from their interferometric millimeter-wave
observations (beamsize of $10\arcsec\times10\arcsec$), the J=1$\rightarrow$0
transition of CO and suggested that the U-shaped morphology was due to the
interactions of multiple outflows, one of which originates with IRAS
22266+6845.  Finally, \citet{lee05} observed L1221 in several molecular lines
(CO, HCO$^+$, N$_2$H$^+$, and CS) with the Berkeley Illinois Millimeter Array
(BIMA) and give 3 mm continuum fluxes for MM1 and MM2 (8 and 4 mJy,
respectively, with beamsize $4.7\arcsec\times3.8\arcsec$), whose positions are
coincident with SMM1 and SMM2.  The continuum fluxes correspond to masses of
0.02 and 0.01 M$_\odot$ for MM1 and MM2, respectively.  \citet{lee05} provide
a detailed model of the kinematics in L1221-MM1.  They see evidence of a
north-south outflow from L1221-MM2, but it is weak and not discussed in
detail. They discuss the presence of a possible binary system in MM1, which
is detected by these \it Spitzer\rm\ observations, and drives an east-west
outflow. The eastern component of this binary is coincident with the position
of IRAS 22266+6845, but \it IRAS\rm\ was unable to resolve the binary pair.
\citet{lee05} did not detect any continuum emission around the west object,
for which they knew the position from these \it Spitzer \rm observations, and
concluded there was very little dust around it.  \citet{lee05} determined that
the observations are best simulated by an infalling, slowly rotating, ringlike
envelope around the the binary system in L1221-MM1.

\section{Observations and Data Reduction} 
 
L1221 was observed with the \it Spitzer Space Telescope\rm\ \citep{werner04},
as part of the Legacy Project ``From Molecular Cores to Planet-forming Disks''
\citep[c2d]{evans03}. L1221 was observed with two instruments on \it
Spitzer\rm: the Infrared Array Camera \citep[IRAC]{fazio04} at 3.6 (IRAC band
1), 4.5 (IRAC band 2), 5.8 (IRAC band 3), and 8.0 $\mu$m (IRAC band 4) and the
Multiband Imaging Photometer for \it Spitzer\rm\ \citep[MIPS]{rieke04} at 24 (MIPS band
1) and 70 $\mu$m (MIPS band 2). The core was observed with IRAC on 19 July
2004 (Program ID (PID) 139, AOR key 0005165312) and with MIPS on 24 September
2004 (PID 139, AOR key 00094287636).
 
The IRAC observations of L1221 consisted of 8 pointings arranged in a 2 column 
by 4 row grid. Each pointing covers an area of about $5\arcmin\times 5\arcmin$ 
for a total area of approximately $10\arcmin \times 20\arcmin$. The area was 
observed with 4 dithers, or individual images offset by approximately 
10\arcsec, with exposure times of 12 seconds each giving a total exposure time 
of 48 seconds. 
 
The MIPS observation fields were smaller, focusing on the central parts of the 
L1221 core. The 24 $\mu$m observations mapped an area of approximately 
$5\arcmin \times 15\arcmin$, or a 1 column by 3 row grid. One cycle (14 
frames) of 3 second exposures was used at 24 $\mu$m to obtain a total exposure 
time of 42 seconds. The 70 $\mu$m camera mapped an area of about $7.5\arcmin 
\times 15\arcmin$ in a $3\times3$ grid.  Each 70 $\mu$m pointing had a total 
exposure time of 90 seconds obtained through three 10 frame cycles of 3 second 
exposures. 
 
The IRAC and MIPS data were processed through the standard \it Spitzer Science
Center \rm\ pipeline (version S13 for IRAC and MIPS) creating Basic Calibrated
Data (BCDs) before undergoing further processing in the c2d team internal
pipeline.  The c2d pipeline improves the BCDs by removing instrumental
artifacts in the data. A more complete description of this part of the data
reduction is available in the c2d data delivery documentation \citep{evans07}.
The improved data are then mosaicked to create a single map from the
observations using the SSC software MOPEX. Sources are extracted and
photometery is performed using a modified version of the DoPHOT software
\citep{schechter93}. The modifications are described in \citet{harvey06}.
  
In addition, the L1221 region was observed with the Very Large Array (VLA) in
C array configuration on the days of 19 July, 2005 and 25 July, 2005 at 3.6 cm
and 6.0 cm.  Observations were centered on SSTc2d J222807.4+690039, which is
called IRS3 in this paper.  The correlator was setup with four IFs at adjacent
frequencies to produce a continuum bandwidth of 172 MHz.  The data were
calibrated and imaged using the standard routines of \it{AIPS++}\rm.  The
observations and data analysis are similar to the continuum observations of
L1014 reported in \citet{shirley07}.

Finally, observations of the N$_2$H$^+$ ($J=1\rightarrow0$) and CS
($J=2\rightarrow1$) were taken with the 14 meter Five College Radio Astronomy
Observatory (FCRAO).  The beam size at these frequencies is about $55\arcsec$,
and the velocity resolution is about 0.07 km/s.  The observed velocity range
for the N$_2$H$^+$ is from -5.1 to -3.7 km/s; the line width is 0.68$\pm$0.04
km/s.  The observed velocity range for CS is -5.9 to -2.9 km/s; the line width
is 1.5$\pm$0.5 km/s.  Multiple pointings were observed to create maps of the
region in the the N$_2$H$^+$ and CS transitions, which are in the 3 mm
atmospheric window. These observations will be described in a separate paper
\citep{devries09}.

\section{Results} 
 
\begin{figure}
  \plotone{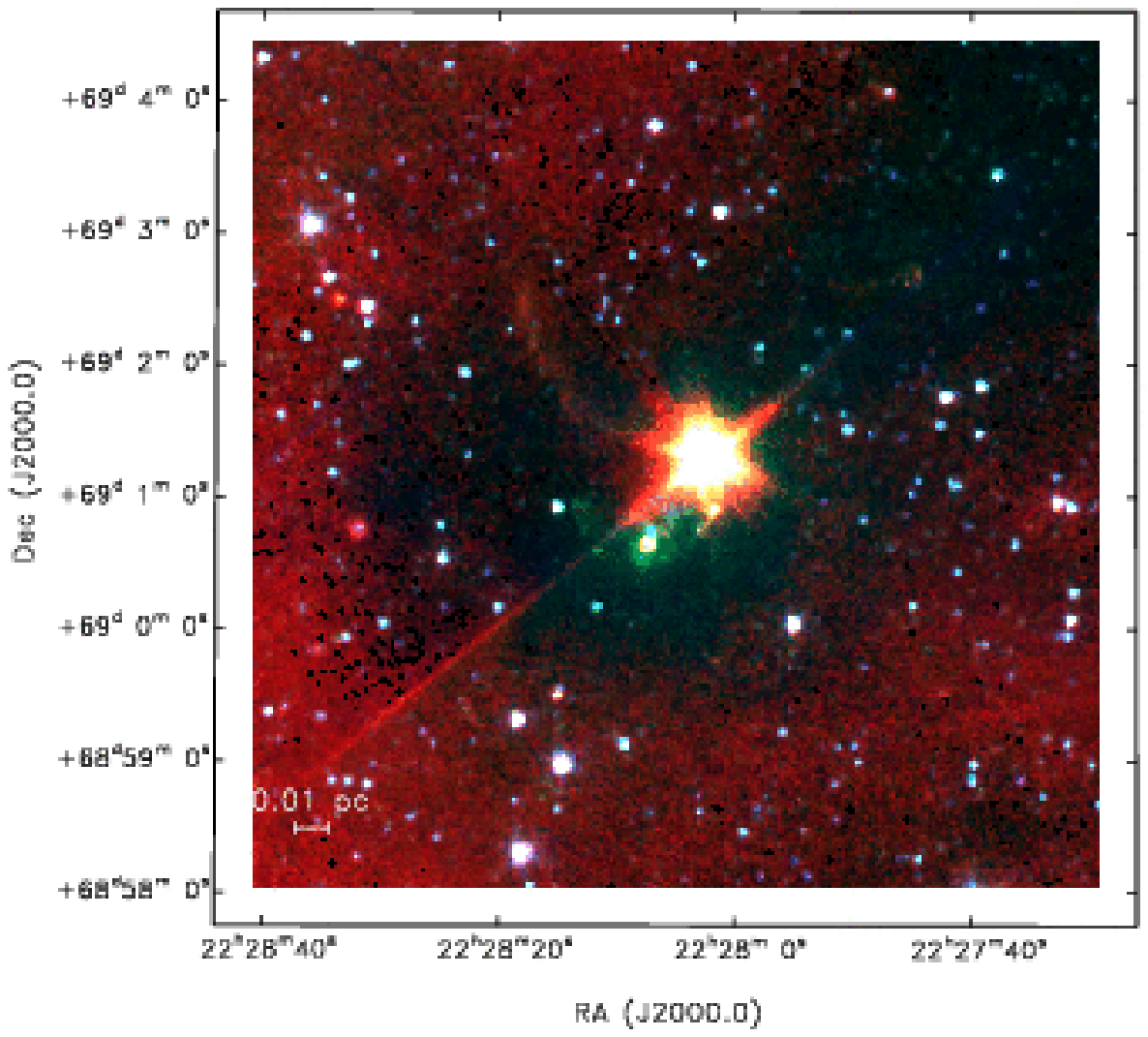} \figcaption{\label{fig-tricolor} Three-color IRAC
    image of L1221. Red corresponds to 8.0 $\mu$m, green is 4.5 $\mu$m, and
    blue is 3.6 $\mu$m. The sources are labeled in Figure~\ref{fig-sixpanel}.
    IRS1/IRS2 is the brightest source; they appear unresolved because the
    source is overexposed in this image. IRS3 is the small, green object
    southeast of IRS1/2. An arc of emission extends from L1221-IRS1 to the
    northeast. L1221-IRS3 shows a signature of a north-south outflow.}
\end{figure}

\begin{figure}
  \plotone{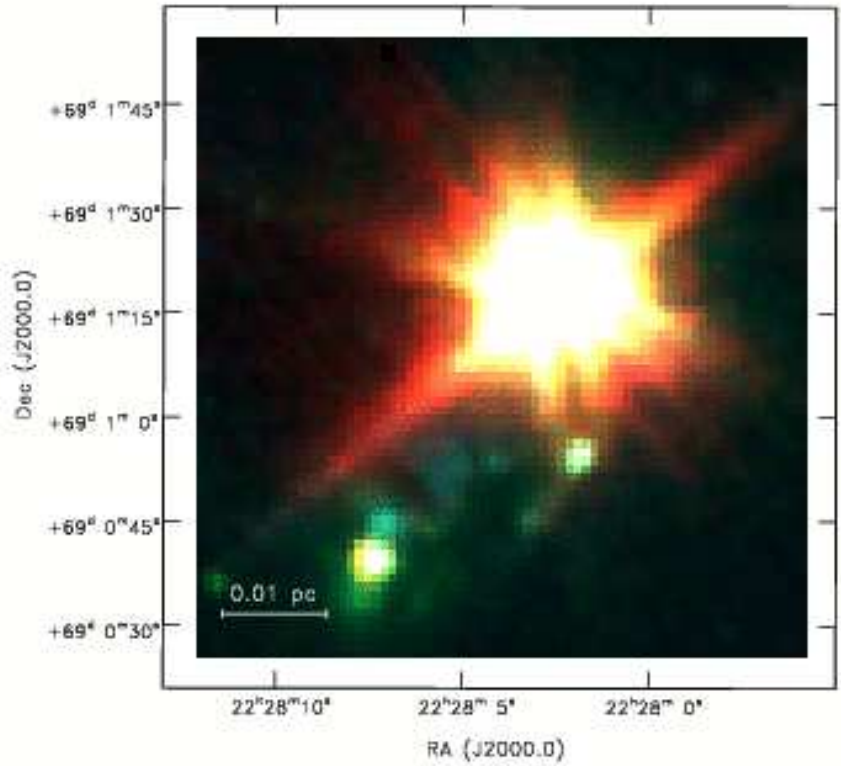} \figcaption{\label{fig-zoom} Three-color IRAC image of
    L1221. The color scheme is the same as in Figure~\ref{fig-tricolor}; the
    image is zoomed in to show detail of the inner region. IRS1 and IRS2 are
    in the brightest source; they appear unresolved because the source is
    overexposed in this image.  IRS3 is the green object southeast of IRS1/2.}
\end{figure}

\begin{figure} 
\centering
 \vspace*{7.8cm}
   \leavevmode
   \includegraphics{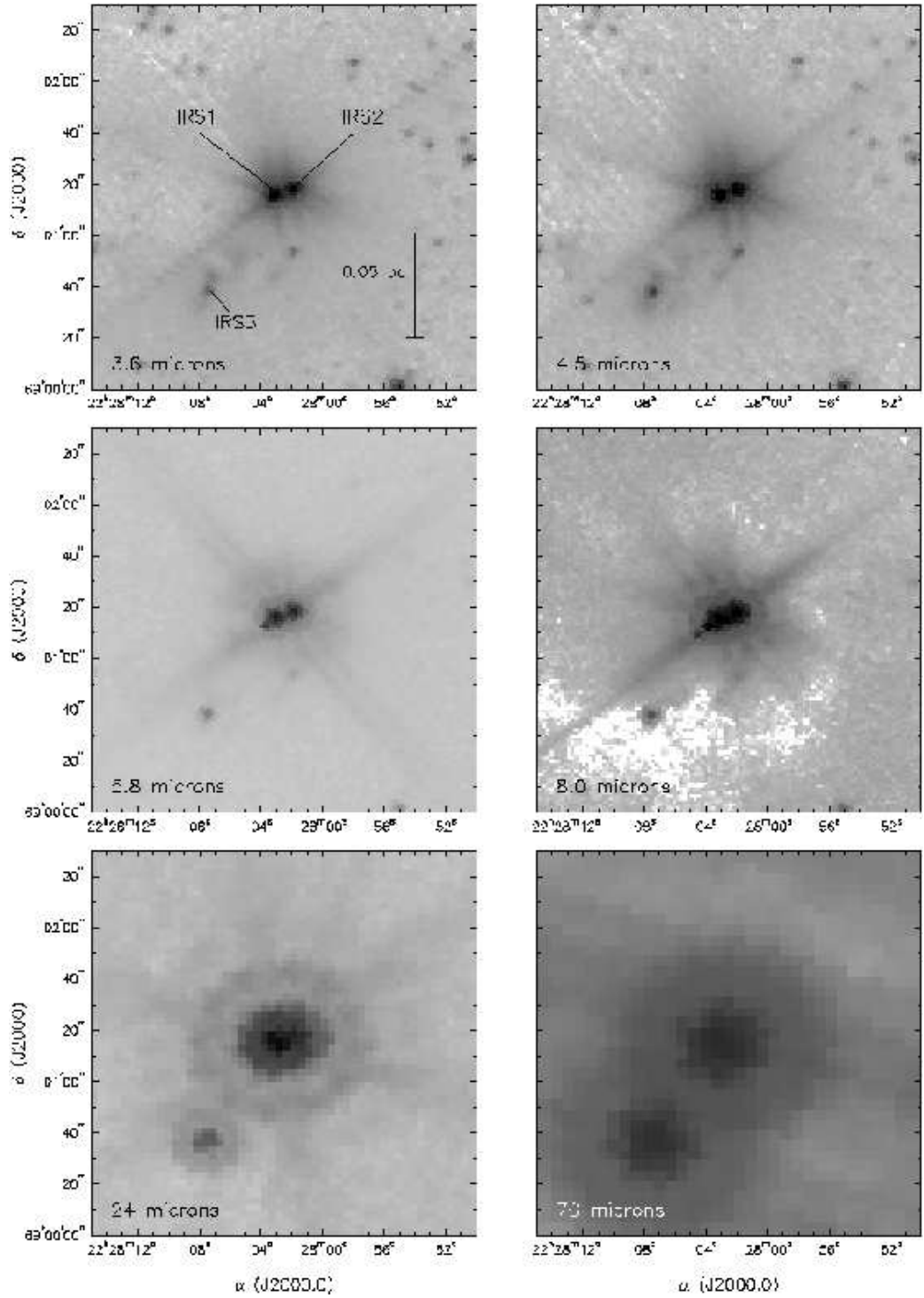}
\vskip 4.in \figcaption{\label{fig-sixpanel} The IRAC and MIPS
    images are shown for L1221.  The maps are shown with logarithmic scaling
    to show IRS1, IRS2, and IRS3.}
\end{figure}

\begin{figure}
  \plotone{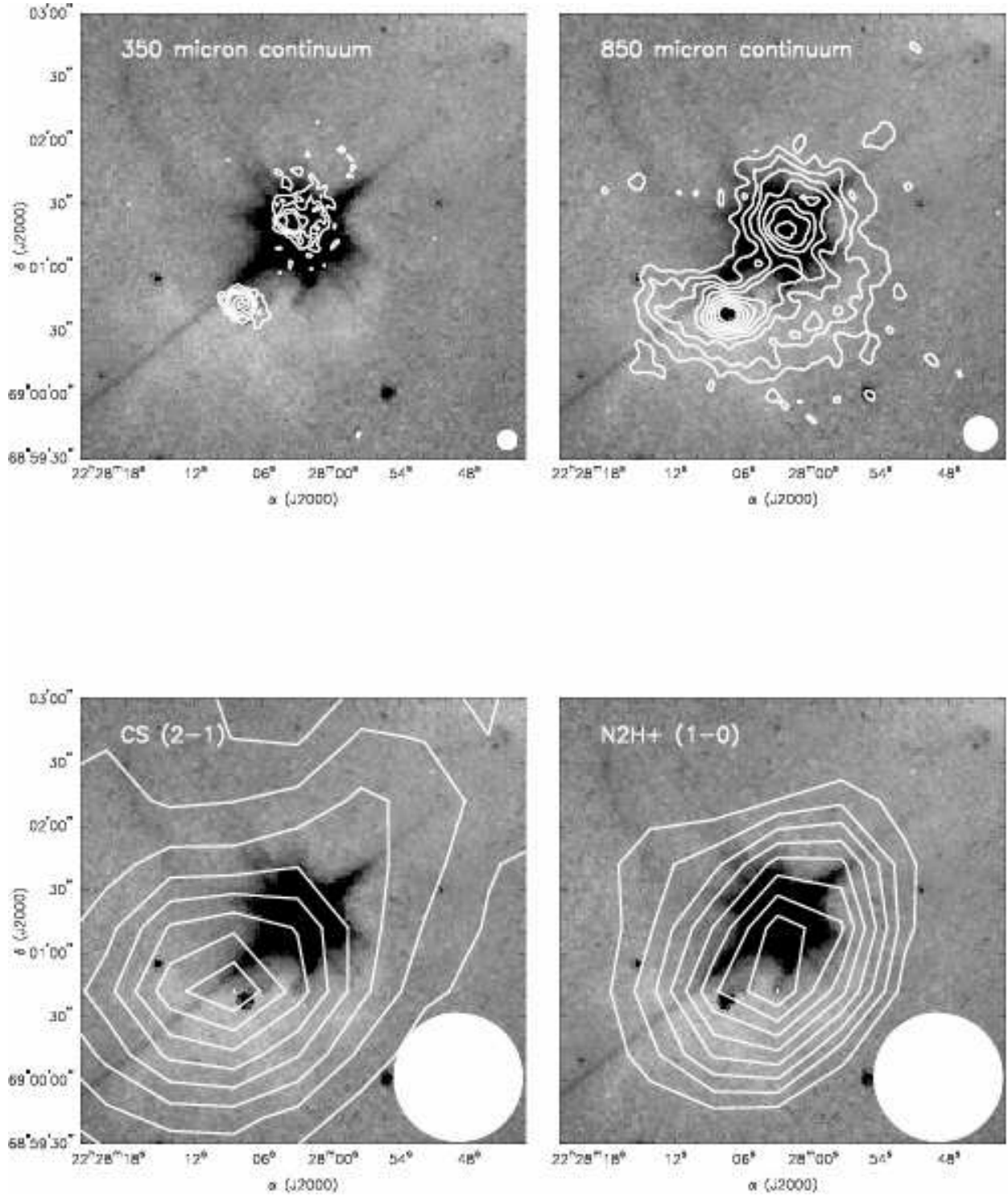} \figcaption{\label{fig-smm} Millimeter
    observations are overlaid, as contours, on the 8 $\mu$m greyscale image.
    The upper left panel shows 350 $\mu$m, upper right is 850 $\mu$m, the
    lower left panel has CS $(J=2\rightarrow1)$, and the lower right panel has
    N$_2$H$^+$ $(J=1\rightarrow0)$.  }
\end{figure}

The images of this region are shown in
Figures~\ref{fig-tricolor},~\ref{fig-zoom},~\ref{fig-sixpanel}, and
~\ref{fig-smm}.  Figure~\ref{fig-tricolor} has the IRAC 3.6, 4.5, and 8.0
$\mu$m observations represented as blue, green, and red. Figure~\ref{fig-zoom}
has the same color scheme but zoomed in to the region of L1221-SMM1 and SMM2.
Three infrared sources--IRS1, IRS2, and IRS3--appear to be associated with
L1221.  IRS1 and IRS2 appear unresolved in Figures~\ref{fig-tricolor} and
\ref{fig-zoom} because the source is overexposed.  The positions of the three
sources are labeled in Figure~\ref{fig-sixpanel}, which show the IRAC and MIPS
data, separately, for the 3 infrared sources.  Finally, Figure~\ref{fig-smm},
has the submillimeter continuum (850 and 350 $\mu$m) and FCRAO molecular line
observations overlaid on the 8.0 $\mu$m greyscale image.  Observed fluxes for
the infrared sources are in Table~\ref{table-sed}.

SSTc2d J222803.0+690117 (hereafter, L1221-IRS1), was detected at all \it
Spitzer \rm bands and was also detected by \it IRAS\rm\ (IRAS 22266+6845) and
\it 2MASS\rm\ (2MASS 22280298+6901166).  L1221-IRS1 was also detected in the
near-infrared by \citet{hodapp94} and catalogued as an outflow driver. This
object has $[4.5]-[8.0] = 1.4$ and $[8.0]-[24] = 4.4$, which qualifies it as a
candidate young stellar object (YSOc) according to the criteria given in
\citet{harvey06}; it is also a YSOc after the scheme in \citet{harvey07}.
Further evidence of its nature include coincidence with dense gas and evidence
that it drives an outflow \citep{lee05}.  The fluxes for L1221-IRS1 (1.25 to
850 $\mu$m, as in Table~\ref{table-sed}) give L$_{bol}$=1.8 L$_\odot$ and
T$_{bol}$=250 K, which classifies it as a Class I object, according to the
scheme developed by \citet{myers93}; these were calculated using the
trapezoidal method of integration. The spectral index, as defined by
\citet{lada87}, is $\alpha =0.81$, which classifies this object as a Class I
protostar \citep{greene94}.  The spectral index is calculated as described in
the final delivery document for the c2d Legacy Program \citep{evans07}.

There is a source 7$\arcsec$ (1750 AU) west of L1221-IRS1, SSTc2d
J22801.8+690119 (hereafter, L1221-IRS2).  IRS1 and IRS2 were not resolved by
\it IRAS\rm; however, the coordinates for the \it IRAS\rm\ source are closer
to IRS1 than IRS2 (4$\arcsec$ versus 17$\arcsec$). The separation between IRS1
and IRS2 is about equal to the resolution of \it 2MASS\rm, so IRS2 is not
included in the \it 2MASS\rm\ catalogs. \citet{hodapp94} detected several
near-infrared sources towards L1221.  The brightest is clearly associated with
L1221-IRS1; they also detect faint emission from IRS2.  The separation between
IRS1 and IRS2 is smaller than the resolution for MIPS at 70 $\mu$m,
17$\arcsec$.  However, the source position for the 70 $\mu$m detection is
coincident with L1221-IRS1.  Therefore, an upper limit of about 300 mJy at 70
$\mu$m can be assumed for L1221-IRS2; a flux larger than this value would
cause a shift in the source position towards L1221-IRS2 (T.  Brooke priv.
comm.).  Additionally, the 24 $\mu$m flux for L1221-IRS2 was not listed in the
c2d source catalogs, so the IRAC source positions were input to MOPEX, which
gave the 24 $\mu$m flux of 480 mJy (Brooke priv.  comm.).  L1221-IRS2 has
$[4.5]-[8.0] = 2.1$ and $[8.0]-[24] = 2.9$, both of which qualify it as a YSOc
\citep{harvey06}; it is also a YSOc after the scheme in \citet{harvey07}. The
bolometric luminosity (from 3.6-24 $\mu$m) is 0.4 L$_\odot$ and T$_{bol}$ =
450 K (Class I).  The spectral index is $\alpha = -0.05$; L1221-IRS2 is
considered a ``flat-spectrum'' protostar by this index \citep{greene94}.  Of
course, the spectral energy distribution is poorly sampled, and these
classifications may change with future observations at different wavelengths.

\it Spitzer\rm\ detected a southeastern source at all wavelengths (SSTc2d
J22807.4+690039), hereafter L1221-IRS3. This object is about 50$\arcsec$
(12500 AU) from L1221-IRS1 and was undetected by \it IRAS\rm\ or \it 2MASS\rm.
Figures~\ref{fig-tricolor} and~\ref{fig-zoom} show 4.5 $\mu$m (green)
emission emanating north and south of IRS3. This emission has been shown to
indicate outflowing material \citep{noriega04}. Indeed, \citet{lee05} detected
a weak, N-S outflow from SMM2. \citet{lee05} proposed that this outflow
probably emanates from IRS3.
 
L1221-IRS3 has $[4.5]-[8.0] = 1.1$ and $[8.0]-[24] = 5$, which qualifies it as
a YSOc; it is also a YSOc after the scheme in \citet{harvey07}. The fluxes
give L$_{bol}$ = 0.8 L$_\odot$ and T$_{bol}$ = 68 K over all observed
wavelengths (3.6-850 $\mu$m), which classifies it as a Class 0 object ($<$70
K). The spectral index is $\alpha = 0.99$ (Class I); the spectral index does
not define the Class 0, so these disparate classifications are not necessarily
inconsistent.

\begin{figure}
  \plotone{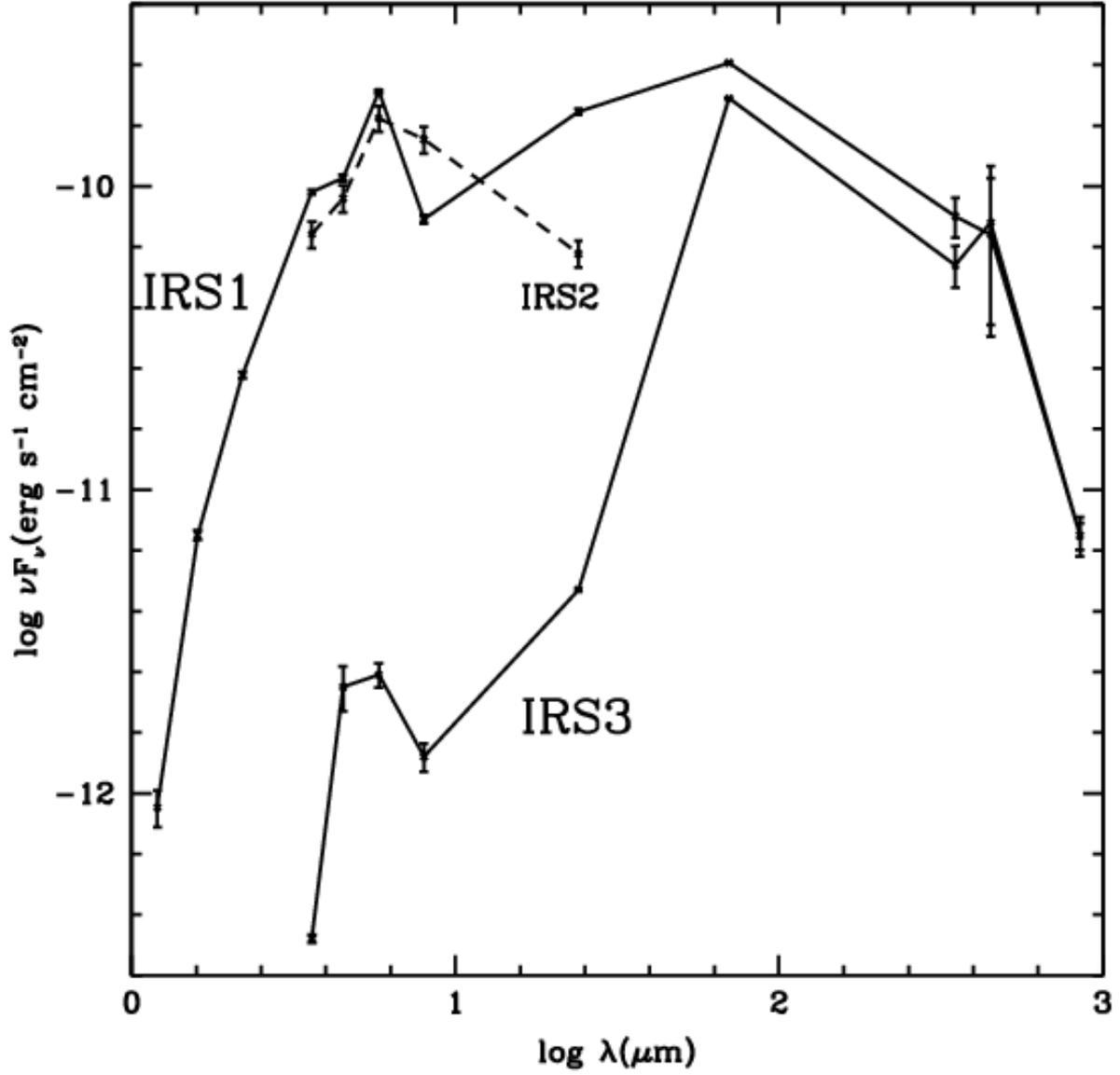} \figcaption{\label{fig-sed} Spectral energy
  distributions for L1221-IRS1, IRS2, and IRS3. IRS1 and IRS3, shown as solid
  lines, were modelled while IRS2 (dashed line) was not.}
\end{figure}

The spectral energy distributions for L1221-IRS1, IRS2, and IRS3 are shown in
Figure~\ref{fig-sed}.  IRS1 shows more emission at shorter wavelengths but is
very similar to IRS3 in the longer wavelengths, suggesting their envelopes,
traced by submillimeter emission, are very similar.
 
L1221-IRS3 was the only source of these three that was detected at either 3.6
or 6.0 cm with the VLA.  This result agrees with the previous VLA image at 3.6
cm of the L1221 region made by \citet{rodriguez98}.  \citet{rodriguez98}
detect an unresolved 3.6 cm continuum source that is coincident with
L1221-IRS3 (22$^h$ 55$^m$ 7.36$^s$ $+69^\circ$ 00$\arcmin$ 39.9$\arcsec$,
J2000.0) with a flux of $210 \pm 20$ $\mu$Jy.  L1221-IRS3 was detected again
in the 2005 VLA observations with at 3.6 cm and 6.0 cm with fluxes of $177 \pm
24$ $\mu$Jy and $192 \pm 27$ $\mu$Jy respectively.  These flux levels are
within a factor of 2 of the centimeter fluxes observed toward L1014-IRS, which
has 3.6 cm and 6.0 cm fluxes of 111$\pm$8 $\mu$Jy and 88$\pm$11 $\mu$Jy,
respectively \citep{shirley07}.  Also, the 3.6 cm flux for L1221 agrees with
the value quoted by \citet{rodriguez98}.  The spectral index between 3.6 cm
and 6.0 cm is $-0.16 \pm 0.54$.  This spectral index is consistent with the
index for optically thin free-free emission, although a steeper negative or
positive spectral index cannot be ruled out.  Since there is strong evidence
from the IRAC images of a molecular outflow from a central protostar at
L1221-IRS3, the most likely explanation for the observed centimeter emission
is shock-ionization from interaction of the protostellar jet from IRS3 with the
surrounding material in the envelope
\citep{curiel87,curiel89,anglada95,shang04}.

Finally, there is a fourth infrared source, which was detected from 3.6-70
$\mu$m.  SSTc2d J222815.1+685930 is an object about 2.2$\arcmin$ southeast
from the L1221 core.  It is included here because it was one of three sources,
including L1221-IRS1 and L1221-IRS3, that was detected at 70 $\mu$m.  The c2d
catalogs list this object as very likely a background galaxy
 based on color-magnitude cutoffs from \it SWIRE\rm\ 
\citep{evans07,harvey07,lonsdale03}.  Fluxes for this source are given in
Table~\ref{table-sed}.

\section{1-Dimensional Models} 
 
Physical models for L1221-IRS1 and L1221-IRS3 were created to best match the
fluxes from 3.6 to 850 $\mu$m; because the SED of L1221-IRS2 is poorly
sampled, there was no attempt to model its emission. The models are
one-dimensional and included a central star, a disk, and envelope. The disk is
included in this one-dimensional model as discussed in \citet{butner94} and
\citet{young05}.  The radiative transfer was calculated using Dusty
\citep{ivezic99}.
 
The model parameters, which were allowed to vary, include the stellar
luminosity (L$_\ast$), the disk luminosity ($L_D$), the stellar photospheric
temperature (T$_\ast$), the inner radius of the envelope (r$_i$), and the
power-law density distribution of the envelope ($p$).
 
The other parameters of the model, which were not changed, include the disk
inner and outer radii (R$_i$ and R$_o$), the power-law surface density
distribution of the disk, the power-law temperature distribution of the disk
(q), properties of dust in the envelope, the outer radius of the envelope
(r$_o$), and the total mass of the envelope (M$_{env}$).
 
For the disk parameters, default values for all except the disk luminosity,
$L_D$, have been adopted. The surface density power law index is $p = 1.5$,
and the temperature power law index is $q = 0.5$. The former is in accordance
with the density structure for a disk in vertical hydrostatic equilibrium
\citep{chiang97}, and the latter is chosen to simulate the effects of flaring
and disk accretion \citep{butner94,kenyon87}. The inner radius of the disk is
set to where the dust is heated to greater than 2000 K and is, presumably,
destroyed. The outer radius is assumed to be 5 AU; this is similar to the disk
radius, based on the centrifugal radius, used for IRAM 04191+1522
\citep{dunham06}.  The effect of the disk outer radius ($R_o$) is discussed in
\S\ref{section-lint}.  The scale of the power-law density distribution was
set by assuming the disk has a mass of 0.005 M$_\odot$, which is similar to
the assumed disk for IRAM 04191+1522 \citep{dunham06}, a source that seems to
be similar to L1221-IRS3.  In fact, the mass of the disk has little effect on
the model. The disk luminosity, $L_D$, arises because of accretion onto the
disk and is a free parameter in these models.  Basically, $L_D$ has to be
tuned in order to match the 24 and 70 $\mu$m observations. As discussed in
\citet{young05}, $L_D$ is not indicative of the mass of the disk; it is simply
set to a particular value so as to match the observations.
 
The 850 $\mu$m flux is least susceptible to geometric effects because the 
envelope is optically thin at this wavelength. Therefore, this flux was used 
to constrain the mass of the envelope.  The fiducial density was set, 
depending on the value for the inner radius, such that the modelled 850 $\mu$m 
flux would match the observed flux. 
 
Because the shape of the envelope's density distribution is not known, a
power-law density distribution, $n(r)=\left( \frac{r}{r_f}\right) ^{-p} $
($r_f = 1000$ AU), with indices of $p=1.5$ or $p=2.0$ is assumed.
\citet{young03} found a median p of 1.8; indeed, most studies have found an
average power-law index to be between 1.5 and 2.0 \citep{shirley02,motte01}.
 
The outer radius of the envelope is chosen to be $r_o=5000$ AU.  The outer
radius is not easily constrained from submillimeter observations.  However,
\citet{wu07} gave the major axes of IRS1 and IRS3 (46$\arcsec$ and
28$\arcsec$; 12000 and 7000 AU) at the 2-$\sigma$ level. These sizes
correspond to outer radii of 6000 and 3500 AU.  Therefore, the outer radius is
set to the average of these values (5000 AU) for both cores.  \citet{wu07}
also warned that SHARC-II is not sensitive to extended emission, so the
reported values potentially underestimate the submillimeter size of the cores.
However, the cores are separated by 50$\arcsec$ in the plane of the sky, which
corresponds to 12500 AU.  Then, it is reasonable to assume the outer radii do
not exceed 6000 AU.

For the dust properties, the opacities calculated by \citet{ossenkopf94} from
the fifth column of Table 1 in their paper (so-called OH5 dust) were used.
These data have been extended to a greater range of wavelengths as described
in \citet{young05}.  The scattering of light by the dust grains is ignored, as
discussed in \citet{young05}.  The impact of this assumption is discussed in \S\ref{section-lint}.

The envelope is heated externally by the interstellar radiation field (ISRF).
As described in \citet{young05}, the ISRF has been attenuated by
the opacity due to \citet{draine84} dust with $A_V = 3$. This simulates the
effect of low density material in the surrounding environs of a star-forming
core.
 
To determine the best-fit model, the reduced $\chi^2$ was calculated over all
\it Spitzer\rm\ wavelengths as such
\begin{equation} 
\tilde{\chi}^2=\frac{1}{k}\sum_{i}\frac{\left[ S_\nu^{obs}(\lambda_i)-S_\nu^{mod}(\lambda_i)\right] ^2 } 
{[\sigma_\nu(\lambda_i)]^2}, 3.6\le\lambda_i\le70 \mu m 
\end{equation} 
The degrees of freedom ($k=n-m$) is the difference between the number
of free parameters ($m=4$) and the number of data points in the SED ($n=9$).

Then, four different grids were examined to find the parameters that give the
lowest ${\tilde{\chi}^2}$ values. These four grids cover the following parameter spaces:
$L_\ast-L_D$, $L_\ast-r_i$, $T_\ast-r_i$, and $T_\ast-L_\ast$, for grids 1
through 4, respectively. 

\begin{figure} 
\centering
 \vspace*{7.8cm}
   \leavevmode
   \includegraphics{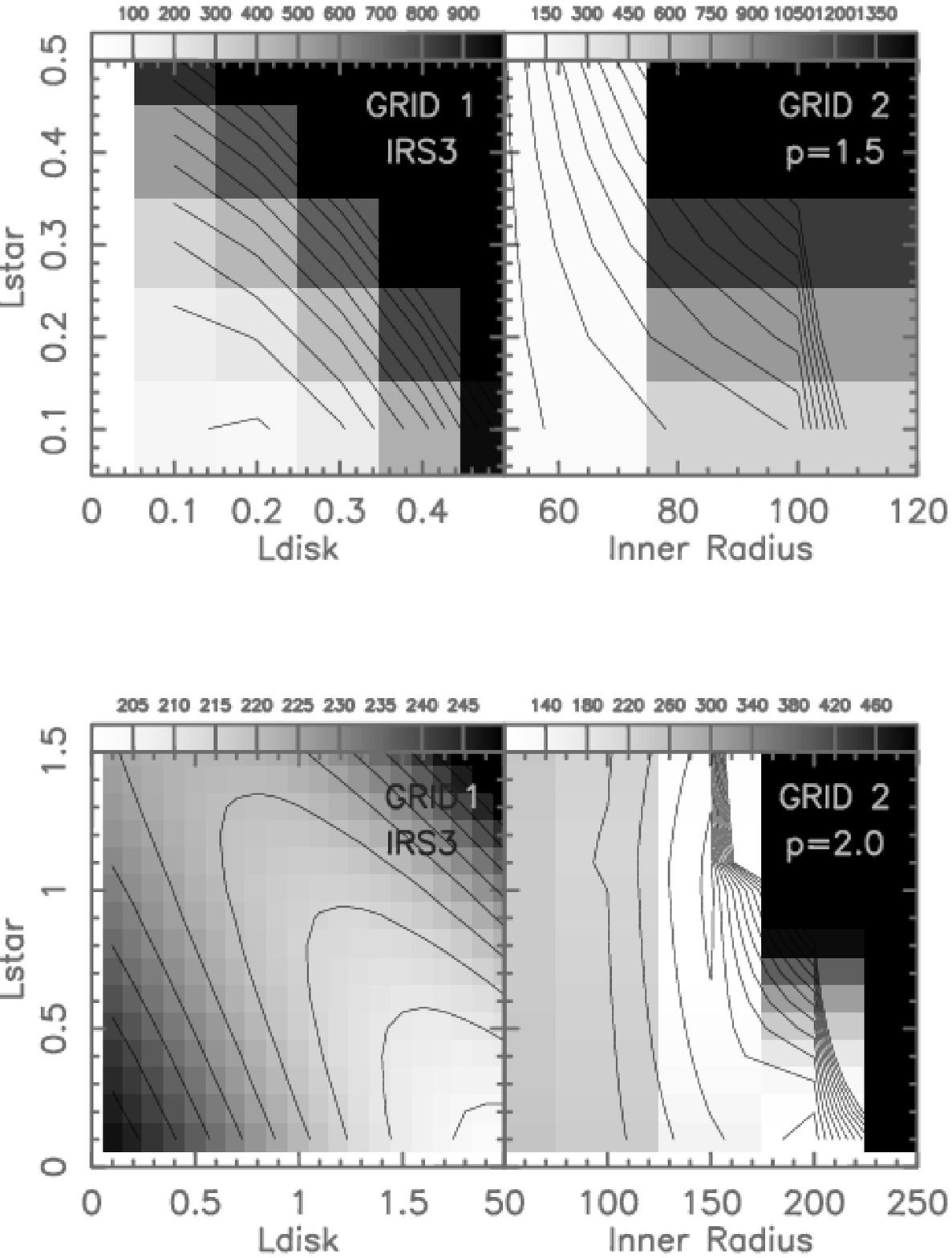}
\vskip 4.in \figcaption{\label{fig-chisq_irs3} The ${\tilde{\chi}^2}$
    values for models of IRS3 with $p=1.5$ and $p=2.0$.  White areas denote
    lower ${\tilde{\chi}^2}$ values. The default values, except when allowed to vary,
    are $L_\ast=0.4$ L$_\odot$, $L_D=0.4$ L$_\odot$, $r_i=100$ AU, and
    $T_\ast=3000$ K.}
 \end{figure}

\begin{figure}
  \plotone{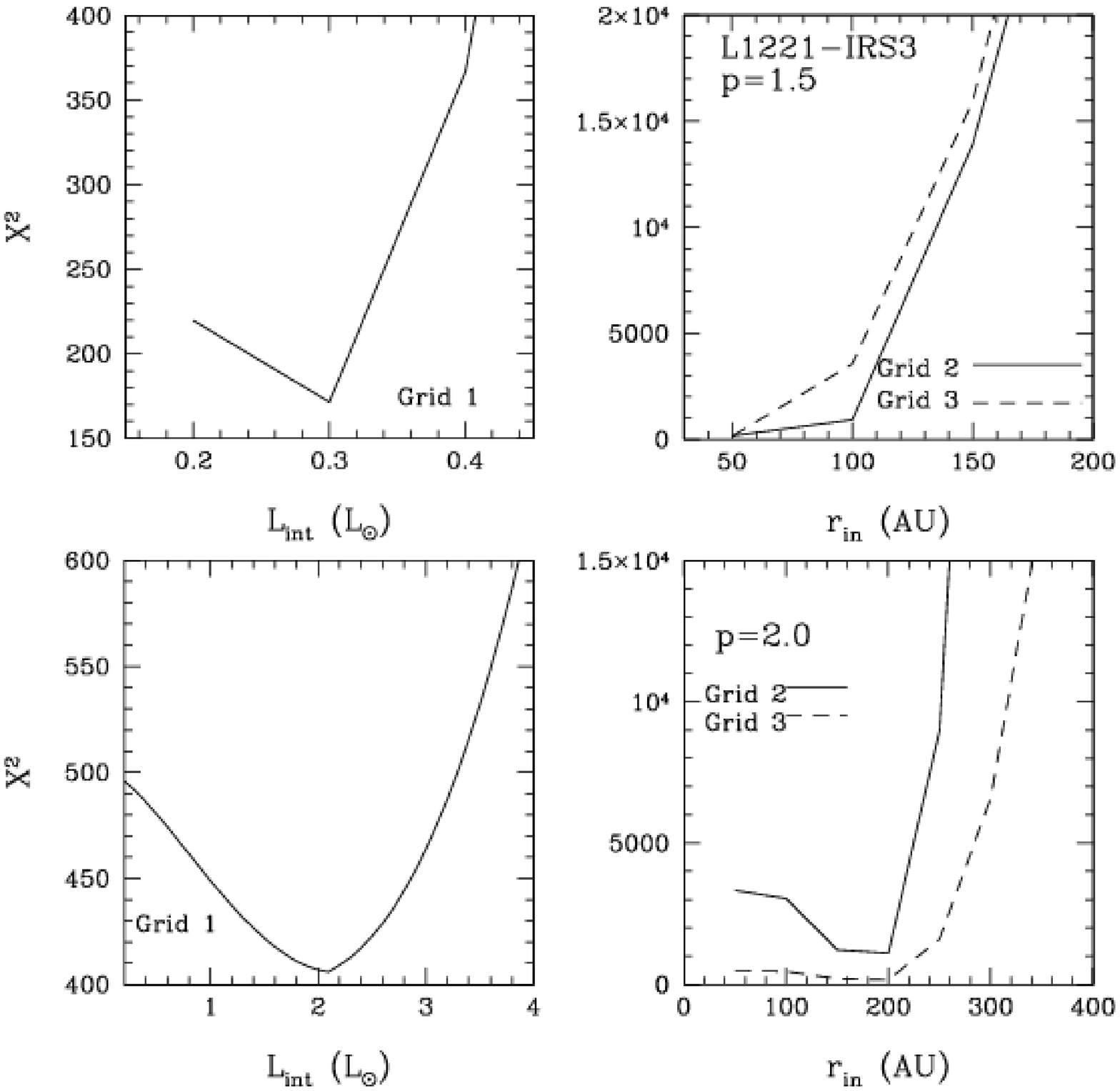} \figcaption{\label{fig-minchisq_irs3} The
    minimum ${\tilde{\chi}^2}$ values plotted from the data shown in
    Figures~\ref{fig-chisq_irs3}.}
\end{figure}

The ${\tilde{\chi}^2}$ values for grids 1 and 2 are plotted as greyscale in
Figure~\ref{fig-chisq_irs1} for L1221-IRS1 and Figure~\ref{fig-chisq_irs3} for
L1221-IRS3.  Grids 3 and 4, which are not shown, pinpoint values for $r_i$ and
$L_\ast$ that are similar to those found in grids 1 and 2.  In addition, grids
3 and 4 probe the effective temperature of the protostar.  These models are
insensitive to stellar temperature. The envelope is optically thick at the
shorter wavelengths, so all of the stellar photospheric radiation is
reprocessed.  The stellar temperature, then, is set at 3000 K, which is
representative for a late-M dwarf.  Because the ${\tilde{\chi}^2}$ plots for
Grids 3 and 4 are redundant, they are not shown, but the minimum
${\tilde{\chi}^2}$ values for several parameters from all grids are plotted in
Figures~\ref{fig-minchisq_irs1} and ~\ref{fig-minchisq_irs3}.
 
\subsection{L1221-IRS1} 
L1221-IRS1 was detected by \it IRAS\rm, \it Spitzer\rm, \it 2MASS\rm, \it
SHARC-II\rm, and \it SCUBA\rm.  However, the near-infrared data are not
well fitted by a one-dimensional model, so the \it 2MASS\rm\ observations are
not used in calculating ${\tilde{\chi}^2}$ (as given by equation 1).
 
L1221-IRS1 has a nearby companion, L1221-IRS2, about 7$\arcsec$ away, which
corresponds to 1750 AU. However, L1221-IRS2 is much less luminous than
L1221-IRS1; it was not resolved at 70 $\mu$m, but there is an upper limit on
the 70 $\mu$m flux of 300 mJy. Additionally, \citet{lee05} detected no
continuum 3 mm emission from L1221-IRS2 and concluded that there was probably
very little dust surrounding this object. \citet{hodapp94} detected emission
from both IRS1 and IRS2, though the emission from IRS2 was very faint.

Because IRS2 is much less luminous, it is not included in the model of IRS1
and its surrounding envelope.  It is, of course, possible that L1221-IRS2 is
not even associated with the core and is simply a background object. However,
it is a YSOc, so its impact as a potential binary companion will be discussed
in section 6.
 
Initially, values for the model parameters that produced a reasonably good fit
to the data were assumed: $r_i = 1000$ AU, $L_\ast = 1$ L$_\odot$, $L_D = 1$
L$_\odot$, and $T_\ast = 3000$ K.; these are the default values for the free
parameters. When allowed to vary, these parameters were set in the following
ranges: $r_i=250 \rightarrow 2500$ AU, $L_\ast=0.5 \rightarrow 5$ L$_\odot$,
$L_D=0.5\rightarrow 5$ L$_\odot$, and $T_\ast=1000\rightarrow 20000$ K.
 
The fiducial density, at $r_f = 1000$ AU, was set such that the total envelope
mass was always a certain value to match the 850 $\mu$m flux (2 Jy). For the
models with $p=1.5$ , the mass is 1.1 M$_\odot$; when $p=2.0$ , $M_{env} =
0.9$ M$_\odot$.  Also, when the envelope has different inner radii, the
fiducial densities must be adjusted to produce these envelope masses. For
inner radii from 250 to 2500 AU, the range of fiducial densities were
$1.8\times 10^6$ to $2.7 \times 10^6$ cm$^{-3}$ for the $p = 1.5$ models and
$2.2\times 10^6$ to $4.2\times 10^6$ cm$^{-3}$ for the $p = 2.0$ models.
 
\begin{figure} 
\centering
 \vspace*{7.8cm}
   \leavevmode
   \includegraphics{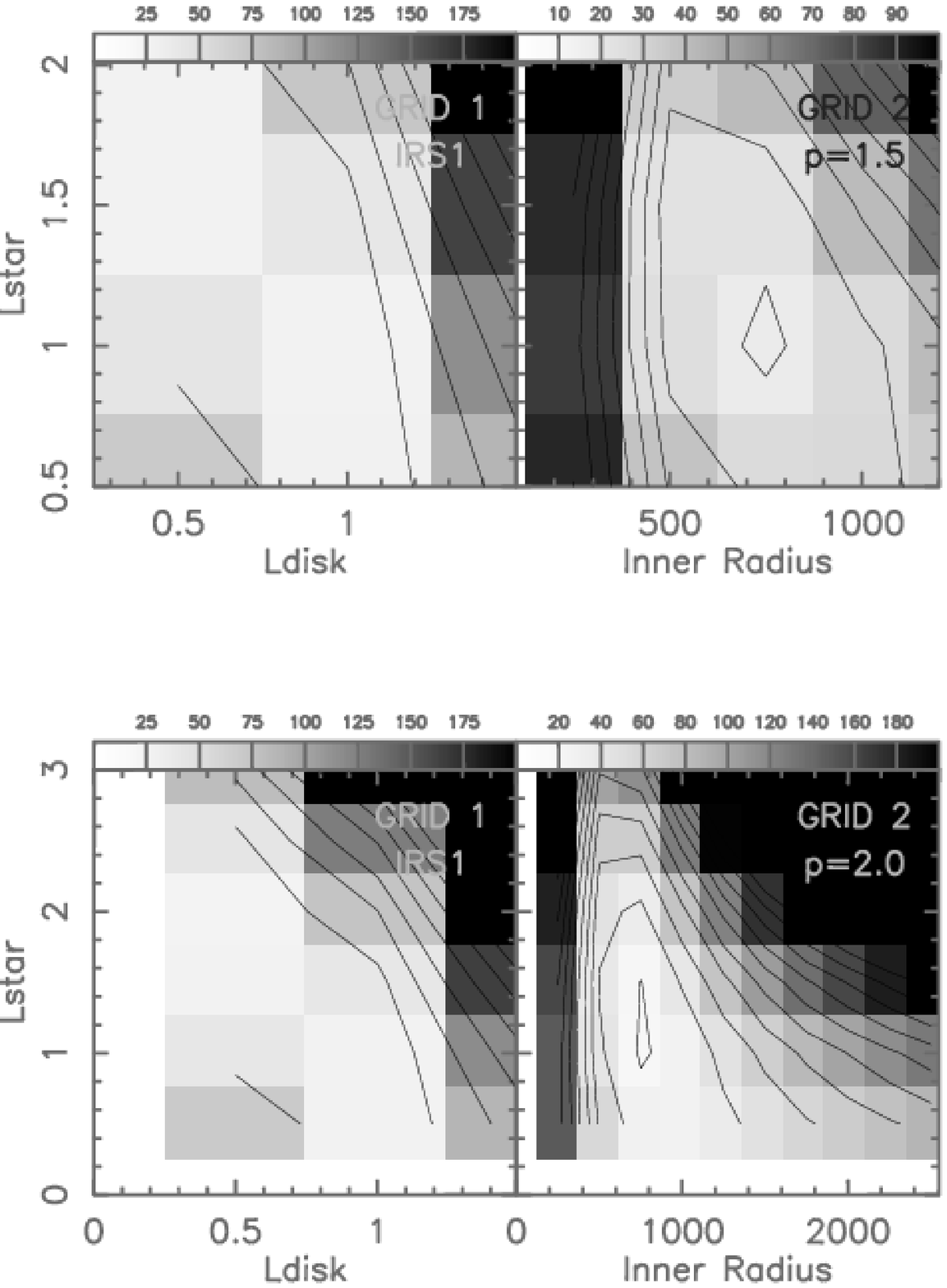}
\vskip 4.in \figcaption{\label{fig-chisq_irs1} The ${\tilde{\chi}^2}$
    values for models of IRS1 with $p=1.5$ and $p=2.0$.  White areas denote
    lower ${\tilde{\chi}^2}$ values. The default values, except when allowed to vary,
    are $L_\ast=1.0$ L$_\odot$, $L_D=1.0$ L$_\odot$, $r_i=1000$ AU, and
    $T_\ast=3000$ K.}
 \end{figure}

Figure~\ref{fig-chisq_irs1} shows greyscale plots with the ${\tilde{\chi}^2}$
values of models with $p = 1.5$ and $p=2.0$.
Grid 1 shows a range of appropriate luminosities for the star and disk.  Grid
2 effectively pinpoints one valid inner radius and stellar luminosity.  
 
\begin{figure}
  \plotone{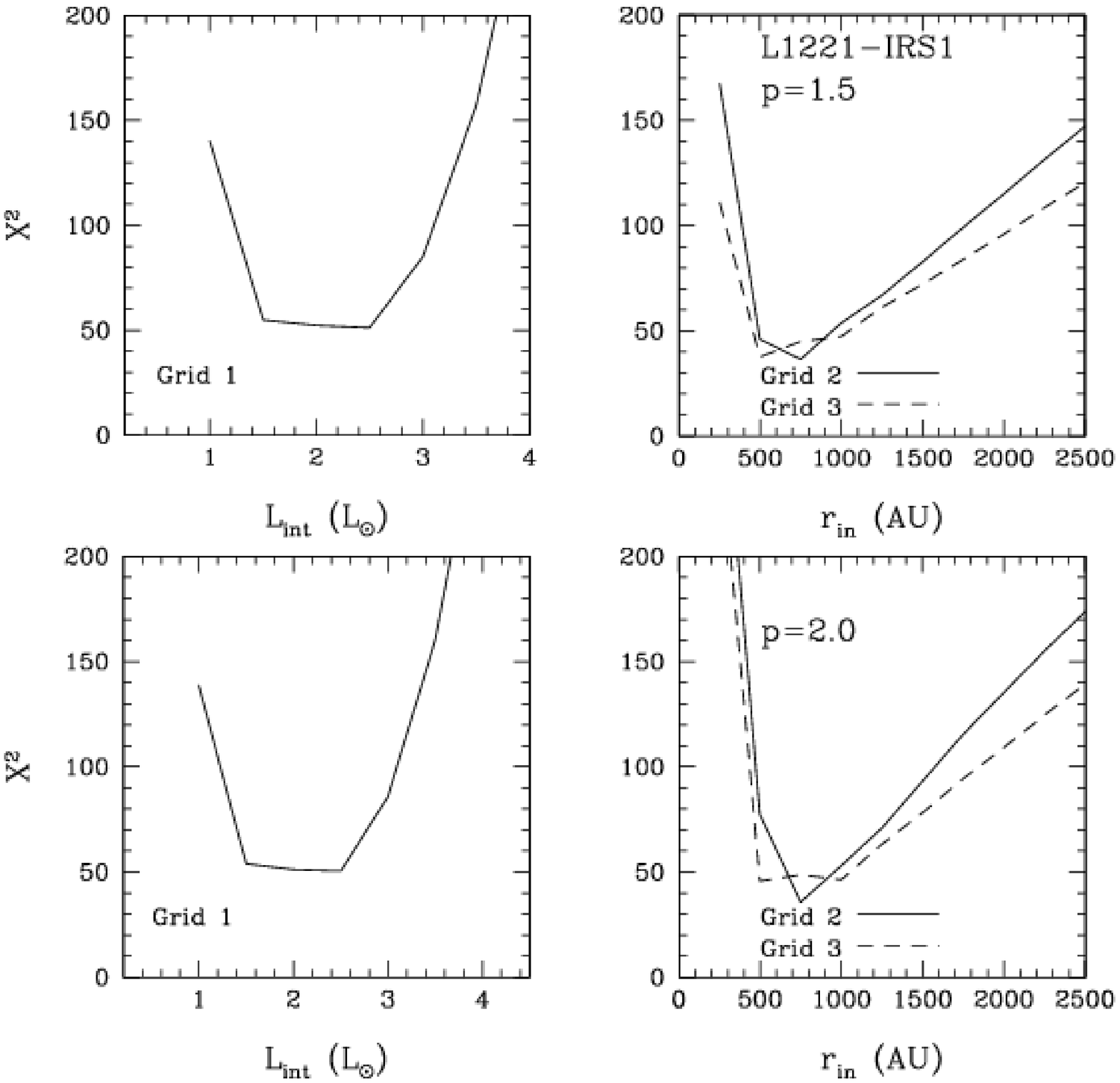} \figcaption{\label{fig-minchisq_irs1} The
    minimum ${\tilde{\chi}^2}$ values plotted from the data shown in
    Figure~\ref{fig-chisq_irs1}.}
\end{figure}

For each set of models, the minimum ${\tilde{\chi}^2}$ values for the total
internal luminosity ($L_{int}$) and $r_i$ are plotted in
Figure~\ref{fig-minchisq_irs1}. These graphs show which of these values gave
the best ${\tilde{\chi}^2}$ and were used in determining the best-fit
parameters.
 
To determine the appropriate model parameters, the inner radius is selected
from Grid 2. Then, the inner radius, which gives the lowest
${\tilde{\chi}^2}$, is taken from Grid 3 (see
Figure~\ref{fig-minchisq_irs1}).  Averaging the values for $r_i$ gives the
best-fit model parameter as reported in Table~\ref{table-bestfit}.  There is
some ambiguity in the knowledge of $L_\ast$ and $L_D$. As shown in Grid 1 of
Figure~\ref{fig-chisq_irs1}, the best-fit stellar luminosity depends upon the
chosen disk luminosity.  Therefore, the sum $L_\ast+L_D$ ($L_{int}$) is
constrained with Grid 1 and equally apportioned for $L_\ast$ and $L_D$.

\begin{figure}
  \plotone{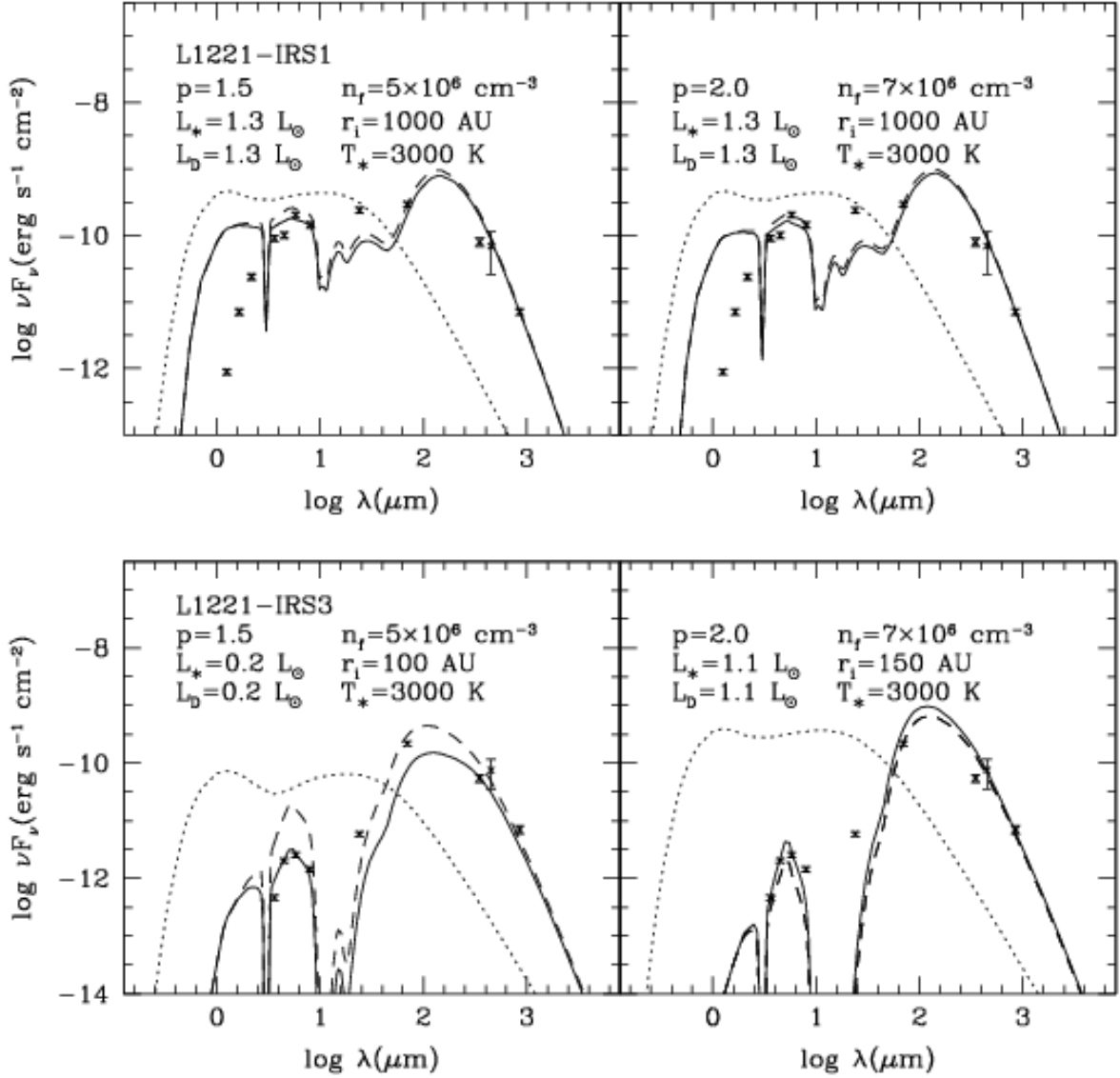} \figcaption{\label{fig-sed_best} The best-fit models
    for differing power-law indices have different parameters as shown in this
    figure for both L1221-IRS1 and L1221-IRS3.  Best-fit parameters are in
    Table~\ref{table-bestfit}.  The dotted line represents the SED of the star
    and disk; the solid line is the SED of the star, disk, and envelope.  The
    dashed line is the model that best-fits only the MIPS observations.  The
    error bars are the observed fluxes.}
\end{figure}

The best-fit parameters are given in Table~\ref{table-bestfit}.  The model with
$p=2.0$ has a slightly better ${\tilde{\chi}^2}$ value; however, both models
have fairly high ${\tilde{\chi}^2}$ values ($\sim$50). For both $p = 1.5$ and
$p=2.0$, the inner radius ($r_i$) is 1000 AU; the total internal luminosity is
2.6 L$_\odot$. The spectral energy distributions for these two models are
shown in the upper panels of Figure~\ref{fig-sed_best}.  The dotted line in
these plots represents the spectral energy distribution (SED) of the star and
disk, excluding the envelope; the solid line is the SED of the star, disk, and
envelope system.  The error bars represent the observed fluxes.
 
The models often did not match the MIPS data, so $\tilde{\chi}^2_{MIPS}$,
which only includes $\lambda_i=$24 and 70 $\mu$m, is used to find the model
that best fits the MIPS data.  $L_{int}$ was held constant while $L_D$ was
varied to optimize $\tilde{\chi}^2_{MIPS}$ while the envelope parameters
($r_i$, $p$, and $n_f$) are the same as in Table~\ref{table-bestfit}.  These
models are represented by the long dashed line in Figure~\ref{fig-sed_best}.
For $p=1.5$, $L_\ast=1.3$ and $L_D=1.9$ L$_\odot$; for $p=2.0$, $L_\ast=1.3$
and $L_D=1.7$ L$_\odot$; the best $\tilde{\chi}^2_{MIPS}$ values were for the
$p=1.5$ model.  In both cases, the 70 $\mu$m observations are better matched,
but the 24 $\mu$m is still underestimated by the model.  As $L_D$ increases,
the 70 $\mu$m flux increases at a greater rate than the 24 $\mu$m flux because
of the higher extinction at 24 $\mu$m.  Therefore, as $L_D$ is increased, the
70 $\mu$m flux is increased, but the 24 $\mu$m remains about the same.  Of
course, the IRAC fluxes are also overestimated by these models, since their
values were not considered in calculating $\tilde{\chi}^2_{MIPS}$. This effect
is prevalent in all four models.  A more realistic and two-dimensional
treatment of the disk is necessary to remove this effect.

\subsection{L1221-IRS3} 
 
L1221-IRS3 is the only infrared source detected in the southern submillimeter
core, L1221-SMM2.  This is the first report of its infrared detection.
Modelling of IRS3 follows closely the process for IRS1.  The same default
values for disk parameters (except $L_D$ and $R_i$), dust opacity (OH5),
power-law density profiles ($p=1.5$ and 2.0), and heating by the ISRF as used
for IRS1 were kept for IRS3.  

Initially, values for the model parameters that produced a reasonably good fit
to the data were assumed: $r_i = 100$ AU, $L_\ast = 0.4$ L$_\odot$, $L_D = 0.4$
L$_\odot$, and $T_\ast = 3000$ K; these are the default values for the free
parameters. When allowed to vary, these parameters were set in the following
ranges: $r_i=50\rightarrow 750$ AU, $L_\ast=0.1\rightarrow 2$ L$_\odot$,
$L_D=0.1\rightarrow 2$ L$_\odot$, and $L_\ast=500\rightarrow 5000$ K.
 
Then, the fiducial density ($n_f$) was set to match the modelled and observed
850 $\mu$m fluxes. The 850 $\mu$m flux is dependent on the dust mass and
temperature. The models with different values for $p$ have differing
temperature distributions. Therefore, one must choose 2 envelope masses that
create an 850 $\mu$m flux that best matches the observations.  For the $p=1.5$
models, the envelope mass is 1.75 M$_\odot$. When $p=2.0$, $M_{env} = 1.25$
M$_\odot$.  The steeper density profile has higher temperatures, so a smaller
dust mass is required to produce the 850 $\mu$m flux.
 
Because the inner radius changes in these models, different fiducial densities
are required to create an envelope with these masses. For inner radii from 50
to 750 AU, the range of fiducial densities (at $r_f = 1000$ AU) were
$2.8\times10^6$ to $2.9\times 10^6$ cm$^3$ for the $p=1.5$ models and
$3.0\times 10^6$ to $3.5\times 10^6$ cm$^3$ for the $p=2.0$ models.
 
The ${\tilde{\chi}^2}$ two-dimensional plots are in
Figure~\ref{fig-chisq_irs3}.  The plots of minimum ${\tilde{\chi}^2}$ are in
Figure~\ref{fig-minchisq_irs3}. Analyzing these data in the same way as for
IRS1, the best-fit parameters were found, which are given in
Table~\ref{table-bestfit}, are as follows. For $p=1.5$ , $r_i=100$ AU, and
$L_{int}=0.4$ L$_\odot$.  For $p=2.0$ , $r_i = 150$ AU and $L_{int}=2.2$
L$_\odot$.  The model with $p=2.0$ has a slightly better ${\tilde{\chi}^2}$
value, though both models have fairly high ${\tilde{\chi}^2}$ values.  The
spectral energy distributions for these two models are shown in the lower
panels of Figure~\ref{fig-sed_best}.

As discussed for IRS1, models considering just $\tilde{\chi}^2_{MIPS}$ were
calculated and are shown by the long-dashed lines in
Figure~\ref{fig-sed_best}. The best-fit models have the same parameters for
the envelope ($r_i$, $p$, and $n_f$) as those in Table~\ref{table-bestfit},
but their disk luminosities are varied.  The new parameters are as follows:
for $p=1.5$, $L_{int}=$0.2 and $L_D=$1.0 L$_\odot$; for $p=2.0$, $L_{int}=$1.1
and $L_D=$0.4 L$_\odot$.  The best $\tilde{\chi}^2_{MIPS}$ values were for the
$p=1.5$ model.

The modelled $L_{int}$ is significantly different than the observed bolometric
luminosity (0.8 L$_\odot$).  This discrepancy highlights the importance of
far-infrared observations.  For the $p=2.0$ model, the luminosity between 75
and 350 $\mu$m is 1.9 L$_\odot$.  However, this spectral range, which is the
peak of the SED, is not sampled with observations, so the modeled L$_{bol}$ is
much larger than the observed L$_{bol}$.

\subsection{Constraint of Internal Luminosity}\label{section-lint}

These models are one-dimensional and insufficient to fully describe the
physical nature of these systems.  One-dimensional models are not equipped to
simulate the effects of a more realistic disk or scattering by dust grains.
These shortcomings cause an uncertainty in the modelling of fluxes at shorter
wavelengths, especially in the near-infrared.  As such, the NIR fluxes have
not been included in the ${\tilde{\chi}^2}$ calculations.  However,
one-dimensional models are useful because they constrain the luminosity of the
internal source.  Either including scattering of light or altering disk
parameters does not affect the conclusions about internal luminosity.

Because dust grains preferentially forward scatter light, the one-dimensional
code, Dusty, is unable to properly account for scattering \citep{young05}.
\citet{dunham09} have created dust opacities that account for scattering by
adding the scattering and absorption opacities. This is not an entirely
reasonable assumption, but \citet{dunham09} have shown that this method offers
good agreement with two-dimensional models, which are able to treat scattering
in a more correct manner.

\begin{figure}
  \plotone{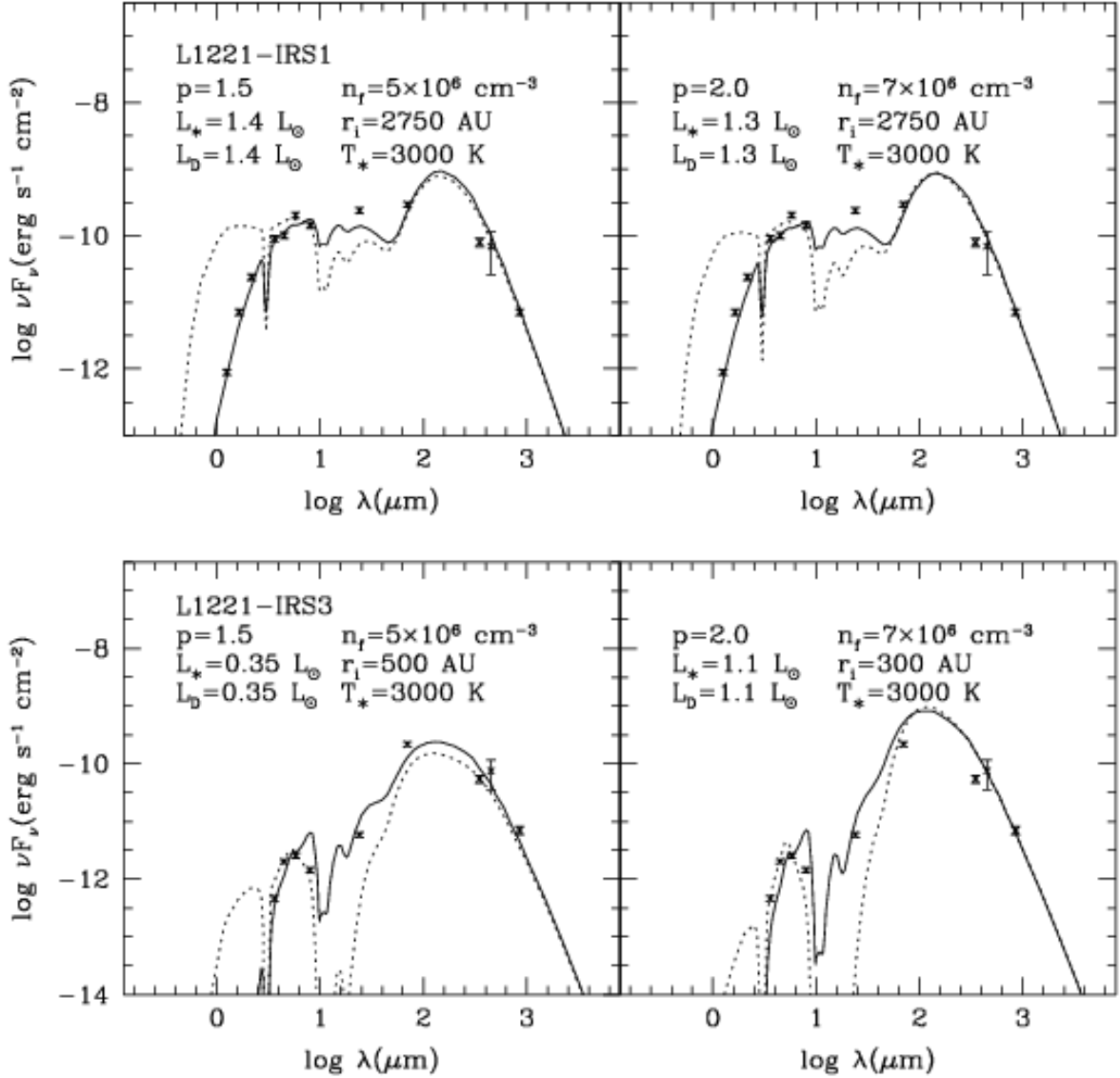} \figcaption{\label{fig-scat} The effects of
  scattering have been included in the SEDs shown in this figure.  The dotted
  line shows the models from Figure~\ref{fig-sed_best} while the solid line
  shows best-fit models when scattering is included.  The parameters for these
  best-fit models are as labeled here.  The internal luminosity is mostly
  independent of the inclusion of scattering; the inner radii increase by a
  factor of 2 or more when scattering is included.
}
\end{figure}

Figure~\ref{fig-scat} shows the best-fit models, whose parameters are given in
Table~\ref{table-bestfit}, as a dotted line.  The solid line in
Figure~\ref{fig-scat} shows the best-fit model when these new opacities are
used (as in \citet{dunham09}).    The new opacities allow for a better fit at
shorter wavelengths, especially in the near-infrared. However, the internal
luminosity for IRS1 and IRS3 are changed very little; the inner radii for both
sources are increased by, at least, a factor of 2.  

\citet{dunham06} concluded, similarly, that the luminosity of the internal
source is well-constrained by one-dimensional models.  The results for
$L_{int}$ of IRS1 and IRS3 are robust, but the inner radii are dependent on a
variety of parameters related to the disk, dust opacities, and other
two-dimensional effects.

\begin{figure}
  \plotone{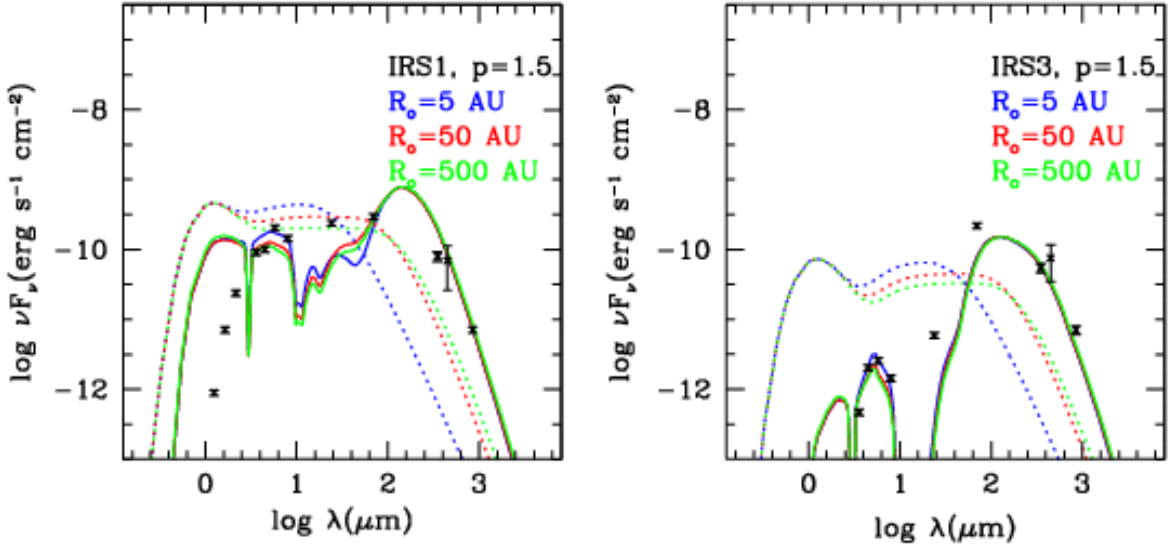} \figcaption{\label{fig-disks} SEDs for IRS1 and IRS3
    ($p=1.5$) in which the disk outer radius has been changed.  The blue shows
    the SED with $R_o=5$ AU, red is $R_o=50$ AU, and green is $R_o=500$ AU.
    Black error bars are the observed data.}
\end{figure}

In addition, the disk is important to the overall fit of the model, but
certain parameters of the disk, such as its outer radius, do not have a large
impact on the best-fit models.  A small disk outer radius (5 AU) has been
adopted for the models of IRS3 and IRS1, but the radius of the disk has little
effect on the modelled SED.  Figure~\ref{fig-disks} has the best-fit models
for IRS1 and IRS3 (with $p=1.5$).  In addition, this figure shows models with
disk outer radii of 50 and 500 AU.  Though the disk spectra (dashed line) are
markedy different, the SED of the protostar, disk, and envelope (solid line)
are all about the same.  Clearly, the disk plays an important role in the
mid-infrared portion of the SED; in fact, these models underestimate the 24
$\mu$m flux.  However, further observations to constrain disk parameters and
higher-dimensional models are needed for more certain conclusions.
Observations with Sofia and Herschel will provide better sampling of the SED,
and submillimeter interferometry will be useful to constrain the disk masses
and sizes in this system.

In comparison, \citet{dunham08} found a relationship between $L_{int}$ and the
flux at 70 $\mu$m; the relationship is found to be reliable within a factor of
$\sim2$.  Based on the 70 $\mu$m flux for IRS1, the derived $L_{int}$ is 1.1
L$_\odot$, while the model suggests $L_{int}=2.6$ L$_\odot$.  The
\citet{dunham08} relationship derives 0.8 L$_\odot$ for IRS3, while the model
gives a range for $L_{int}$ from 0.4 to 2.2 L$_\odot$, depending on the
density profile for the envelope.  Given the uncertainty in the
\citet{dunham08} relationship, the modeled luminosities are consistent with
those derived from the 70 $\mu$m flux.

In conclusion, results for the internal luminosity are fairly robust.  They
are not affected by changes either in the disk or the effects due to
scattering, even though these changes can drastically affect the near-infrared
portion of the overall spectrum.   

\section{Discussion and Conclusions} 
The L1221 star-forming region presents an example of stars forming differently
in very similar environments. L1221-SMM1 and SMM2 are, presumably, both part
of the same region and are separated, in the plane of the sky, by
approximately 50$\arcsec$. Their 850 $\mu$m fluxes are identical, so their
total masses are also similar, though their submillimeter sizes and, possibly,
density profiles, are different.  Also, the two cores appear to be a part of
the same molecular core, as in Figure~\ref{fig-smm}, and dark cloud
\citep{lynds62}. These two cores, because of their close proximity
($50\arcsec= 12500$ AU), are likely to be affected by the same large-scale
dynamical effects due to nearby stars and astronomical events; they are also
likely to have been born out of a very similar makeup of materials. In short,
L1221-SMM1 and SMM2 should provide similar environments for the formation of
stars.

L1221 is not unique.  A number of cores have been found to be home to multiple
protostars in different stages of formation.  \citet{duchene07} found
multiplicity rates of 30-50\% among protostellar systems with separations of
up to 1400 AU; the protostars included Class I and flat-spectrum sources in
several molecular clouds.  IRAM 04191+1422 has several similarities with
L1221, as well.  \citet{andre99} discovered this very-low luminosity object
(VeLLO) \citep{dunham06} about 1\arcmin\ (8400 AU) from IRAS 04191+1523, a
Class I protostar.  Like L1221-IRS1, IRAS 04191+1523 also has two sources,
unresolved by \it IRAS\rm, that are 6.5\arcsec\ apart.  These sources were
resolved by IRAC and 2MASS but not MIPS or SCUBA \citep{dunham06}.  This
star-forming region is similar to L1221 in that it appears to have several
protostars at different stages of evolution.  L1251B is another star-forming
region that is home to different classes of protostars including four Class
O/I stars \citep{lee06}.  Finally, \citet{yun95} found two near-infrared
sources in the star-forming core CB230.  These supposed protostars are
separated by 11\arcsec.  Other examples of pairs include BHR 71
\citep{bourke01} and CG30 \citep{chen08}.  These cores are examples of sources
similar to the L1221 star-forming region in that they are all forming multiple
protostars that are, presumably, at differing stages of evolution.

IRS1 and IRS3, the dominant infrared sources in SMM1 and SMM2, are clearly not
as similar as their host cores. First, their luminosities are different.
L1221-IRS1 has an internal luminosity, including the disk and stellar
components, of 2.6 L$_\odot$, while L1221-IRS3 has $L_{int} = 2.2$ or 0.4
L$_\odot$, depending on the density profile.  The observed infrared luminosity
(with $\lambda\leq70$ $\mu$m) for IRS1 is 1.6 L$_\odot$ and for IRS3 is 0.4
L$_\odot$.

The luminosity is the result of accretion onto the disk and star. Since each
core likely has similar temperature and turbulence, the accretion rates for
IRS1 and IRS3 are probably very similar.  Because accretion luminosity is
directly proportional to the protostar's mass, IRS1 is potentially about 4
times more massive than IRS3.

However, IRS3, for the two best-fit models, has a range of possible
luminosities from 0.4 to 2.2 L$_\odot$. Additional observations in the
wavelength range from 70 to 350 $\mu$m (such as with Herschel) would easily
discern the appropriate internal luminosity since the models are quite
different in this portion of the SED.

Current submillimeter (450 and 850 $\mu$m) maps, made with the scan map
technique on SCUBA, have poor signal-to-noise ratios and are insufficient to
determine the shape of the density profile for these cores.  More sensitive
observations of the extended submillimeter emission are needed to determine
the density profile of the envelope \citep{shirley02}.

The envelopes' inner radii of the L1221-IRS3 and L1221-IRS1 models are
distinctly different.  L1221-IRS3 requires $r_i = 100$ or 150 AU; L1221-IRS1
needs a much larger inner radius (1000 AU). Two scenarios can explain this
disparity.  First, L1221-IRS3 might be at a much earlier stage in its
collapse, and the inner radius has not expanded as predicted by
\citet{terebey84}.  Another alternative is that the binary possibly in
L1221-SMM1 (IRS1 and IRS2) could have evacuated the cavity of some fraction of
the material. The separation of this binary is $7\arcsec$, which corresponds
to 1750 AU.  \citet{jorgensen05} found a similarly large inner radius for IRAS
16293-2422 and reported that the radius of the inner cavity was comparable to
the centrifugal radius.  \citet{lee05} reported a ``hole'' in the N$_2$H$^+$
emission, which could suggest a cleared out cavity in the center of the core.
However, this could also be a chemical effect as discussed in
\citet{jorgensen04a} and as observed by \citet{jorgensen04b}.

For IRAS 16293-2422, \citet{jorgensen05} suggested a two-dimensional model
that used a much smaller inner radius when outflow cavities were instituted
and when the protostar was viewed down the outflow cavity.  They found that
the mid-infrared fluxes were greatly increased when observed down the outflow
cavity.  A similar scenario might apply for L1221 IRS1 and IRS3.  IRS1 might
be observed pole-on while IRS3 is edge-on.  However, because of the bipolar
nature of IRS1 \citep{lee05}, this is unlikely.

The chief results of these models include these conclusions about the
luminosities and envelope inner radii.  The luminosity of each core is easily
constrained with a one-dimensional model because most of the lumonisity is
emitted in the far-infrared, which is largely unaffected by the central star's
near-infrared spectrum; the one-dimensional model is sufficient also because
the re-radiated emission is thin.  Given the same data, even a more complex
two-dimensional model should reach the same conclusions regarding the
luminosity of IRS1 and IRS3.  Prediction of the envelope's inner radii, on the
other hand, is a less robust conclusion.  The central protostar's spectrum
does have a large impact on the inner radius required to match the data.
However, the models do suggest that the inner radius of IRS1 is much larger
than the inner radius of IRS3.  These results still hold even when the disk is
altered or scattering is included in the model.
 
Finally, IRS1 and IRS2, the potentially binary companion to IRS1, show
different SEDs.  IRS1 is detected at submillimeter wavelengths, while IRS2
shows no clear evidence of long-wavelength emission.  Possibly, IRS2 is more
evolved than IRS1 and has stopped accreting material from the surrounding
envelope.  Perhaps, IRS1 and IRS2 exhibit such different properties because
IRS1 is dominating the accretion.  In fact, this might be common in binaries
that are close enough; initially, one gets bigger and then dominates the
accretion for the duration of their formation.

\section{Acknowledgements} 
 
Support for this work, part of the \it Spitzer \rm\ Legacy Science Program,
was provided by NASA through contract 1224608 issued by the Jet Propulsion
Laboratory, California Institute of Technology, under NASA contract 1407, and
the Smithsonian Astrophysical Observatory, under NASA contract 1279198.  This
work was also supported by NASA grants NAG5-10488 and NNX07AJ72G.  KEY was
supported by NASA under Grant No. NGT5-50401 issued through the Office of
Space Science.  CHY thanks the Louisiana Board of Regents, BoRSF, under
agreement NASA/LEQSF(2001-2005)-LaSPACE and NASA/LaSPACE under grant
NGT5-40115 for support during this project.

\clearpage

\clearpage

\clearpage

\clearpage

\clearpage

\clearpage
\begin{deluxetable}{llrlccccccccc}
\tabletypesize{\scriptsize}
\tablecolumns{3}
\tablecaption{Spectral Energy Distribution\label{table-sed}}
\tablewidth{0pt} 
\tablehead{
\colhead{Source Name}                  &
\colhead{Lambda}                           &
\colhead{F$_\nu$}                          &
\colhead{Aperture}                          \\
\colhead{}                             &
\colhead{$\mu$m}                        &
\colhead{mJy}                         &
\colhead{arcsec}

}
\startdata 

SSTc2d J222803.0+690117    & 1.25 &   0.374$\pm$0.037    & 2.5      \\       
(L1221-IRS1)               & 1.65 &   3.91$\pm$0.39     &  2.5    \\       
L$_{bol}^{obs}$=1.8 L$_\odot$    & 2.17 &   17.3$\pm$1.7 & 2.5     \\       
T$_{bol}^{obs}$=250 K      & 3.6  &   109$\pm$11      &   1.7$^a$   \\       
$\alpha=0.81$              & 4.5  &   149$\pm$15      &   1.7$^a$   \\       
                           & 5.8  &   392$\pm$39      &   1.9$^a$   \\       
                           & 8.0  &   387$\pm$39      &   2.0$^a$   \\       
                           & 24   &   1940$\pm$187       &  5.7$^a$    \\     
                           & 70   &   6940$\pm$641       &  17$^a$    \\
                           & 350  &   9300$\pm$1400        &  40    \\
                           & 450  &   10400$\pm$6500      &  40    \\ 
                           & 850  &   2000$\pm$200        &  40    \\
\tableline
SSTc2d J222801.8+690119    & 3.6  &   83.6$\pm$ 8.4      &  1.7$^a$    \\       
(L1221-IRS2)               & 4.5  &   137$\pm$ 14      &    1.7$^a$  \\       
L$_{bol}^{obs}$=0.4 L$_\odot$ & 5.8 & 324$\pm$32      &     1.9$^a$  \\       
T$_{bol}^{obs}$=450 K      & 8.0  &   381$\pm$38      &     2.0$^a$ \\
$\alpha=-0.05$              & 24   &   368$\pm$35     &      5.7$^a$ \\
                           & 70   &   $<300$    &           17$^a$ \\
\tableline
SSTc2d J222807.4+690039    & 3.6  & 0.567$\pm$0.060       &  1.7$^a$    \\      
(L1221-IRS3)               & 4.5  &   3.1 $\pm$0.3     &     1.7$^a$ \\        
L$_{bol}^{obs}$=0.8 L$_\odot$ & 5.8 & 4.96$\pm$0.50      &   1.9$^a$   \\       
T$_{bol}^{obs}$=68 K       & 8.0  &   3.84$\pm$0.4     &     2.0$^a$  \\        
$\alpha=0.99$              & 24   &   47.5$\pm$4.4     &     5.7$^a$  \\
                           & 70   &   5080$\pm$469       &   17$^a$   \\  
                           & 350  &   6400$\pm$1000     &   40   \\
                           & 450  &   11300$\pm$6100   &   40   \\
                           & 850  &   2000$\pm$300   &     40 \\
\tableline
SSTc2d J222815.1+685930    & 3.6  &   0.208$\pm$0.02     &   1.7$^a$   \\      
                           & 4.5  &   0.284$\pm$0.03     &   1.7$^a$   \\        
L$_{bol}^{obs}$=0.003 L$_\odot$ & 5.8 & 0.38$\pm$0.04      & 1.9$^a$     \\       
T$_{bol}^{obs}$=187 K      & 8.0  &   0.613$\pm$0.06    &    2.0$^a$  \\        
$\alpha=0.25$              & 24   &   2.13$\pm$0.27    &      5.7$^a$   \\ 
                           & 70   &   24.1$\pm$3.8  &      17$^a$ \\   

%
%
%

\enddata 
\tablenotetext{a}{FWHM of the \it{Spitzer} point-spread profile.}

\end{deluxetable}

\begin{deluxetable}{llrlccccccccc}
\tabletypesize{\scriptsize}
\tablecolumns{3}
\tablecaption{Model Parameters\label{table-bestfit}}
\tablewidth{0pt} 
\tablehead{
\colhead{}                  &
\colhead{Parameter}                           &
\colhead{Accepted Value}                          
}
\startdata 
    
L1221-IRS1                 & $p$    &   1.5    \\       
                           & $n_f$  &   5$\times10^6$ cm$^{-3}$ \\       
                           & $r_i$  &   1000 AU \\  
                           & $L_{int}$ &  2.6 L$_\odot$ \\       
                           & $T_\ast$ &  3000 K   \\       
L1221-IRS1                 & $p$    &   2.0    \\       
                           & $n_f$  &   7$\times10^6$ cm$^{-3}$ \\       
                           & $r_i$  &    1000 AU \\       
                           & $L_{int}$ &  2.6 L$_\odot$ \\       
                           & $T_\ast$ &  3000 K   \\
       
L1221-IRS3                 & $p$    &   1.5    \\       
                           & $n_f$  &   5$\times10^6$ cm$^{-3}$ \\       
                           & $r_i$  &   100 AU \\       
                           & $L_{int}$ &  0.4 L$_\odot$ \\       
                           & $T_\ast$ &  3000 K   \\       
L1221-IRS3                 & $p$    &   2.0    \\       
                           & $n_f$  &   7$\times10^6$ cm$^{-3}$ \\       
                           & $r_i$  &    150 AU \\       
                           & $L_{int}$ &  2.2 L$_\odot$ \\       
                           & $T_\ast$ &  3000 K   \\       
                  
\enddata                                              
\end{deluxetable}

\end{document}